\documentclass[twocolumn,aps,prl,superscriptaddress]{revtex4-1}
\usepackage{graphicx}
\usepackage{dcolumn}
\usepackage{bm}
\usepackage{color}
\usepackage{ulem}
\usepackage[english]{babel}
\begin{document}

\title{Optical Rogue Waves in integrable turbulence}  
  
\author{Pierre Walczak}
\affiliation{Laboratoire de Physique des Lasers, Atomes et Molecules,
  UMR-CNRS 8523,  Universit\'e de Lille, France} 
\author{St\'ephane Randoux} 
\affiliation{Laboratoire de
  Physique des Lasers, Atomes et Molecules, UMR-CNRS 8523,
  Universit\'e de Lille, France} 
\author{Pierre Suret}
\email[Corresponding author : ]{Pierre.Suret@univ-lille1.fr}
\affiliation{Laboratoire de Physique des Lasers, Atomes et Molecules,
  UMR-CNRS 8523,  Universit\'e de Lille, France}

\begin{abstract}
  We report optical fiber experiments allowing to investigate integrable
  turbulence in the focusing regime of the one dimensional nonlinear
  Schr\"odinger  equation  (1D-NLSE). Our experiments are very similar
  in their principle to water tank experiments with random initial 
  conditions (see M. Onorato {\it et al.} Phys. Rev. E {\bf 70} 067302 
 (2004)).  Using an original optical sampling setup, we measure precisely 
 the probability density function (PDF) of optical power of partially
 coherent waves rapidly fluctuating with time. The PDF is found to 
 evolve from the normal law to a strong heavy-tailed distribution, 
 thus revealing the formation of rogue waves in integrable turbulence. 
 Numerical simulations of 1D-NLSE with stochastic initial conditions 
 reproduce quantitatively the experiments. Our investigations suggest 
 that the statistical features experimentally observed rely on the 
 stochastic generation of coherent analytic solutions of 1D-NLSE such 
 as Peregrine solitons.  
  \end{abstract}

\maketitle

The field of nonlinear optics has recently grown as a favorable
laboratory to investigate both statistical properties of nonlinear
random waves and  hydrodynamic-like phenomena
\cite{Turitsyna:13,Fatome:14,Picozzi:14}. In particular, 
several recent works point out analogies between hydrodynamics and  
nonlinear fiber optics in the observation of supercontinuum generation
\cite{Chabchoub:13}, undular bores \cite{Fatome:14}, optical
turbulence,  laminar-turbulent transition \cite{Turitsyna:13} or 
oceanographic rogue waves (RW) \cite{Solli:07, Dudley:14}.
 
Rogue waves (RW), also called freak waves, are  extremely large
amplitude waves  occurring more frequently than expected from  
the normal law \cite{Onorato:01,Onorato:13}. Since the pioneering
work of Solli {\it   et al} in 2007 \cite{Solli:07}, optical RW 
have been  studied in various contexts such as supercontinuum 
generation in fibers \cite{Solli:07, Solli:08,   Erkintalo:09, Kibler:09}, 
optical cavities \cite{Montina:09}, semiconductor lasers\cite{Bonatto:11}, 
mode-locked fiber lasers\cite{Lecaplain:12}, laser filamentation
 \cite{Kasparian:09} and Raman fiber amplifiers or lasers
 \cite{Hammani:08, Randoux:12}.

As  stressed out in the recent review \cite{Dudley:14}, there
is no obvious analogy between most of the optical experiments
on extreme events and oceanography. However a direct correspondence between
nonlinear optics and hydrodynamics is provided by the 
one-dimensional nonlinear Schr\"odinger equation (1D-NLSE)  (see Eq.
(1)) that describes various wave systems \cite{Dudley:10,
  Dudley:14}.  In particular, the focusing 1D-NLSE describes at leading order the physics of deep-water wave trains and it plays a central role in the study of RW \cite{Onorato:01,Dudley:10, Onorato:13,  Akhmediev:13, Dudley:14}. 

Modulational instability (MI) is
believed to be a fundamental mechanism for the formation of RW
\cite{Onorato:04,Onorato:13}. Moreover, analytical solutions of the integrable 1D-NLSE  such a
Akhmediev breathers (AB), Peregrine solitons  or  Kuznetsov-Ma
solitons (KMs) are now considered as possible prototypes of RW
\cite{Akhmediev:13,Kibler:10,Chabchoub:11,Kibler:12}. These coherent
structures have been generated from very specific, carefully-designed
{\it coherent} initial conditions in optical fiber experiments  
\cite{Kibler:10,Kibler:12, Frisquet:13}.

On the contrary, oceanic RW emerge from the interplay of {\it
incoherent waves} in turbulent systems. The occurrence of RW in wave
turbulence has been theoretically studied in optics
\cite{Hammani:10,Kibler:11} and in hydrodynamics \cite{Janssen:03}.
In hydrodynamical experiments made with one-dimensional water tanks, 
non gaussian statistics of the wave height has been found to emerge 
from random initial conditions \cite{Onorato:04,Onorato:05}.

The appropriate theoretical framework combining  a statistical
approach of {\it random waves} together with the  1D-NLSE  is {\it
  integrable turbulence}. This emerging fundamental field of research
recently introduced by Zakharov relies on the analysis of complex
phenomena found in nonlinear random waves systems described by an
integrable equation \cite{Zakharov:09,Zakharov:13,Pelinovsky:13, Agafontsev:14c,  Randoux:14}. The mechanisms found in 
integrable turbulence are of a profoundly different nature 
than those found in conventional turbulence
\cite{Zakharov:09, Picozzi:14, Suret:11}. A very recent study of the
formation of RW in integrable turbulence surprisingly shows that MI
leads to stationary statistical state that is not characterized 
by a high occurence probability  of extreme events \cite{Agafontsev:14c}. 

Nonlinear fiber optics is a promising field for the investigation of integrable
turbulence because  optical  tabletop ``model experiments''
precisely described by the 1D-NLSE can be performed  
\cite{Kibler:10,Kibler:12, Frisquet:13, Randoux:14}.  For instance, the statistical
properties of {\it slowly} fluctuating random waves  in {\it defocusing} integrable turbulence have been recently studied \cite{Randoux:14}. As MI easily broadens optical spectrum beyond the bandwidth of standard photodetectors, the experimental study of the {\it focusing} integrable turbulence is very challenging. Despite the numerous works devoted  to optical RW, {\it the  generation of extreme events from purely  stochastic initial conditions in focusing 1D-NLSE model experiments} remains  a crucial and open question \cite{Akhmediev:09b,Akhmediev:13,Agafontsev:14c, Dudley:14}.\\

In this letter, we address the fundamental problem of the propagation of random
waves described by the focusing 1D-NLSE
\cite{Akhmediev:09b,Zakharov:13, Agafontsev:14,Agafontsev:14c}. We
implement an optical fiber experiment conceptually analogous to the water
tank experiment described in \cite{Onorato:04}.  Using an original
setup to overcome  bandwidth limitations of usual detectors, we 
 evidence strong distorsion of the statisticcs of nonlinear random
 light characterizing the occurrence of optical rogue waves in integrable turbulence.\\

Response times of conventional detectors are usually slower than the typical
time scale characterizing power fluctuations of incoherent optical waves. Since the work of Solli 
{\it et al.} spectral filters are
therefore often used to reveal extreme events in time-domain
experiments \cite{Solli:07,Erkintalo:09, Randoux:12}. In addition to
these filtering techniques, shot-to-shot spectrum fluctuations 
can be evidenced with a dispersive Fourier transform measurement 
in experiments of pulsed supercontinuum generation 
\cite{Solli:07,Jalali:10, Wetzel:12,Goda:13}.  Measuring  in
 {\it an accurate and a well-calibrated } way the probability
 density function (PDF) characterizing
 temporal fluctuations of the power of {\it random light} is 
still a challenging task in the field of nonlinear statistical optics.\\

In order to break through currently-existing detection limitations, we have
developed an original setup which allows the precise measurement of
statistics of random light rapidly fluctuating with time.  Inspired by
the time-resolved fluorescence upconversion experiments 
\cite{Mahr:75} and by the optical sampling (OS)
oscilloscope \cite{Duguay:68}, the principle of our method is based
on asynchronous OS (see Fig. 1.a).

Our experimental setup is schematically shown in Fig. 1.b.
A ``continuous'' wave (cw) Ytterbium fiber laser (IPG-YLR series)
emitting a linearly polarized partially coherent wave  at
$\lambda_S=1064$nm is used as a random light source. This cw laser
emits numerous (typically $10^4$) uncorrelated longitudinal modes. 
The partially coherent wave under
investigation is called the ``signal''. Blue pulses are generated at
a wavelength $\lambda=457$nm by sum-frequency generation (SFG) between 
the signal at $\lambda_S=1064$nm and short ``pump'' pulses having  a central
wavelength  $\lambda_P=800$nm. SFG is achieved in a
$5\times5\times8$mm BBO crystal. Non collinear Type I phase matching 
is achieved with an external angle of $10^o$ between the pump and the
signal. 

 \begin{figure}[h]
\includegraphics[width=7cm]{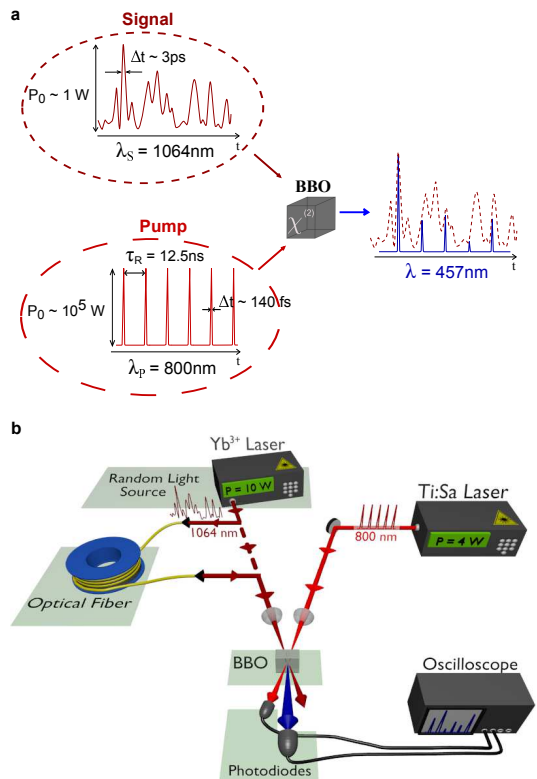}
\caption{ 
  {\bf Measurement of the statistics of random light}
{\bf a.  Principle}. The optical sampling of the partially-coherent wave
  fluctuating with time (the signal) is achieved from sum-frequency
  generation (SFG).  Blue pulses are generated at $\lambda=457$nm from
  the interaction of the signal with femtosecond pump pulses inside a $\chi^{(2)}$ crystal.
 Each SFG blue pulse  has a peak power proportionnal to the  initial signal power. {\bf b. Setup} The $140$fs pump pulses are
  emitted by  mode-locked laser at $\lambda_p=800$nm. The partially coherent
  wave is emitted by a Ytterbium fiber laser at $\lambda_s=1064$nm. We measure 
  statistics of partially-coherent light from the SFG process
  either directly at the output of the laser or after propagation
  inside an optical fiber. Pump pulses and SFG  pulses  are
  simultaneously recorded with  an oscilloscope
}
\end{figure}

The $140$fs-long pump pulses are emitted by a
mode-locked Ti:Sa laser (Coherent Cameleon ultra II) with a repetition
rate of $80$MHz. The maximum output power of the fiber laser 
is much weaker than the peak power ($\simeq 4.10^5$W) of the  pump
pulses. The pump pulses remain therefore undepleted  and the peak
powers of SFG pulses  are proportional to the instantaneous optical
powers $P=P_{(\lambda_s=1064nm)}$ carried by the signal \cite{Boyd}.
The variations of SFG pulses power  (solid red line in Fig. 2.a) 
can be seen as snapshots  of the fluctuations of the optical power $P$ carried
by the signal. We compute the PDF  of $P$
from the statistical distribution  of the  peak powers of SFG pulses
(red line in Fig 2.b).  

The short blue pulses are observed by using a highly sensitive photodiode
(MenloSystem FPD310-FV)  having  a gain of $\simeq 10^4$  and a rise
time of $0.7$ns. We record the output of  the photodiode  with a
fast oscilloscope (Lecroy WaveRunner 104MXi-A, bandwidth 1GHz, 10GS/s).   
We have carefully used the
photodiode in a linear regime without any saturation effect. The peak voltage is proportional 
to the energy of the corresponding optical pulse. A second photodiode 
is used to record pump pulses with a high signal-to-noise 
ratio. This provides a synchronization signal permitting to identify 
the maxima of SFG pulses. The normalized  PDF of the signal is
computed from  an ensemble of  approximately $16$ millions  measurements of 
SFG peak powers.\\

We first measure the PDF at the output of the laser. In all
experiments presented in this letter, the mean output power of the
Ytterbium laser is fixed at $<P>=10$W. At this operating point, the
statistics of the partially coherent wave follows the normal law.
Indeed, as plotted in Fig. 2.b, the PDF of the normalized power
$P/<P>$ is very close to the exponential function (see Fig. 2.b). It
is important to note that the red line on Fig 2.b is not a fitted
exponential function but it represents the exact 
normalized  $\text{PDF}[P/<P>]=\exp(-P/<P>)$.
To the best of our knowledge, PDF of so rapidly fluctuating optical
signals has never been {\it quantitatively} compared to the
normalized exponantial distribution. 

We now use the output of the laser as a random source and we launch
the partially coherent signal into an optical fiber in the focusing
regime. The fiber  is a 15m-long highly nonlinear photonic
crystal fiber (provided by Draka France company) having an anomalous
dispersion at $1064$nm. The fiber maintains the polarization of light
and single transverse mode propagation is also achieved. A random light
wave with a mean power $<P>=600$mW is launched into the fiber.

Experiments have been carefully designed to be very well described by
1D-NLSE. In particular, the signal wavelength $\lambda_s=1064$nm is
far from the zero-dispersion wavelength  $\lambda_0$$\simeq 970$nm.
Moreover the optical spectral widths (see Fig. 2.c)  remain
sufficiently narrow to neglect stimulated Raman scattering (SRS) and
high-order dispersion effects.  The linear losses  experienced by
optical fields  in single pass in the fibers are very negligible.
These total losses are around $0.3\%$  in the fiber  
(loss coefficient of $8$dB/km).\\

\begin{figure}[h]
\includegraphics[width=8.cm]{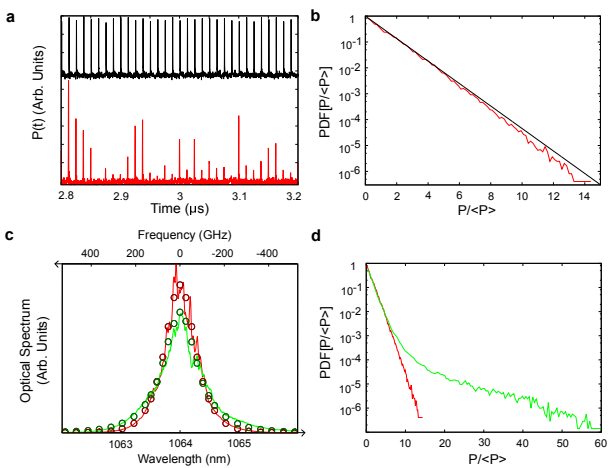}
\caption{ {\bf Experiments} {\bf a.} Pump pulses (black line) and
  samples of fiber laser fluctuations (SFG pulses, red line). {\bf b.}
  Statistics of fluctuations at the output of the fiber laser.
  Probability Density Function (PDF) of normalized optical power
  $P/<P>$ (red  line). The PDF is computed from SFG peak powers plotted in Fig. 2.a.
  Normalized exponential distribution
  $\text{PDF}[P/<P>]=\exp(-P/<P>)$ (black line) {\bf c. and d. Focusing
    propagation} Optical power spectra (c.) and  PDF (d.) of the
  partially coherent wave at  the input (red line) and output (green
  line) of the fiber. Experiments are plotted with solid line (c. and d.) and
  numerical simulations are plotted with circle (c).}
\end{figure}

The measurement of the statistics of the optical power after
propagation of the partially coherent field in the fiber reveals the
occurrence of numerous extreme events (RW). The comparison between the
initial PDF (see red line in Fig 2.d) and the output PDF (see green curve
in Fig. 2.d) shows an impressive change in the statistical
distribution of optical power. The initial field follows the normal
law and its PDF is an exponential function whereas the output PDF of
optical power exhibits a strong heavy-tail.  The probability of
occurrence of very high powers fluctuations (more than 10 times
greater than the mean power)  is much larger than the probability
defined by the normal law. As an example, a fluctuation with a power
greater than fifty times the mean power almost never occurs in the
initial random gaussian field (one  intense fluctuation every
$10^{10}$ seconds) whereas it occurs every $10^{-6}$ second at the
output of the nonlinear fiber.

We have performed numerical simulations of the 1D-NLSE 

\begin{equation}
  \label{eq:NLS1D}
  i\frac{\partial \psi}{\partial z}=\frac{\beta_2}{2}\frac{\partial^2
    \psi}{\partial t^2}-\gamma|\psi|^2\psi
\end{equation}

with parameters corresponding to the experiments. At
$\lambda_s=1064$nm  the  group velocity dispersion coefficient  of
the fiber is $\beta_2=-20ps^2/km$. The effective Kerr coefficient 
is $\gamma=50$W$^{-1}$km$^{-1}$.

Numerical simulations are performed  by discretizing a temporal window 
of $618$ ps with a set of $8192$ points and by using a pseudo spectral 
splitstep-based method. Mean optical power spectra and PDFs are computed 
from Monte Carlo simulations performed over an ensemble of $4000$ realizations
of the initial random process. For each realization 
of the random initial condition, the
initial field  is computed the Fourier space with
the random phase procedure (\cite{Nazarenko, Suret:11, Barviau:06}) :
$\widetilde{\psi}(\omega)=\sqrt{n(\omega)}\,\exp(i \phi_\omega)$ where
$\phi_\omega$ is a white delta-correlated random process and $n(\omega)=n_0 \,
sech(\omega/\Delta\omega)$ is the initial optical spectrum.  The width 
$\Delta \omega= 2 \pi \times 63$GHz of the power spectrum of the wave 
used as initial condition is obtained from a fit of power spectra 
experimentally recorded. \\

\begin{figure}[h]
\includegraphics[width=8cm]{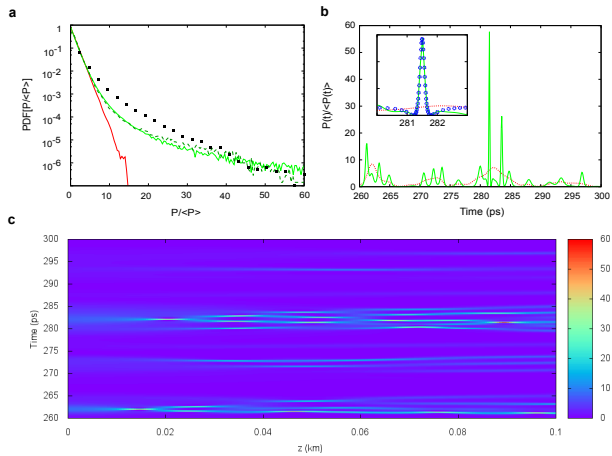}
\caption{ {\bf Numerical simulations.}  {\bf a.} PDF of
  optical power at a propagation length $z=0$m (solid red line),
  $z=15$m (solid green line) and $z=100$m (stationary state, black squares).  Experimental PDF ($z=15$m) of Fig. 2.d is  displayed for  comparison   (dashed green line).  {\bf b. and c.} Typical temporal optical
 power fluctuations along the nonlinear propagations.  Green line in c. :
fluctuations at $z=90$m.  Inset in b. : enlarged view of an intense
power fluctuation superimposed on the profile of a Peregrine soliton (blue circle) determined from a best fit procedure.}
\end{figure}

Optical spectra  (see circles in Fig.2.c) and PDFs (see solid green line
in Fig. 3.a) computed from the
numerical integration of the 1D-NLSE are in
quantitative and remarkable agreement with experiments. The comparison
between experiments and simulations prove that statistical distributions 
found in experiments are very well described by the integrable 1D-NLSE.\\

Numerical simulations of 1D-NLSE allow us to explore the statistics 
of the wave system at longer propagation distances. This is not 
feasible in the experiment because other effects such as stimulated Raman
scattering play a non-negligible role and strongly break the integrability 
of the wave system. With the parameters corresponding to our experiments, 
numerical simulations show that the spectrum and the PDF 
reach a stationary state for a fiber length $L\simeq 100$m. 
The stationary state of integrable turbulence is here characterized 
by a strongly heavy-tailed PDF (see black squares in Fig 3.a). 

In our experiments, the frequency at which the
gain of the MI process is maximum ($275$GHz) lies within the
 spectrum of the initial incoherent wave (see Fig. 2.c). On the
 contrary the author of ref. \cite{Agafontsev:14c} study numerically 
the nonlinear stages of the MI of the condensate. In this case 
the stationary PDF
 of the wave amplitude is a Rayleigh distribution, which means that
 the probability of RW formation is the same as in a linear wave
 system.  It is not surprising in integrable 
systems that the
 nature of the stationary state depends on the initial conditions.
 However further theoretical investigations are needed to fully
 characterize this remarkable difference between two stationary 
states of integrable turbulence.

Numerical simulations  give  some insight into the mechanisms
underlying the breakdown of the gaussian statistics observed 
in our experiments. In the focusing propagation, solitons on 
finite background such as AB, Peregrine solitons or KMs having 
a short duration and a high power seem to
emerge on the top of the highest fluctuations  (see Fig. 3.b). 
The inset of Fig. 3.b evidences an example of the close similarity 
between the shape of an intense and isolated 
power fluctuation and the analytical form of a Peregrine soliton.\\

 The results presented in this letter do not only provide fundamental
 results on rogue waves in integrable turbulence but they also demonstrate
 a powerful method to study the statistics of optical fields rapidly
 fluctuating with time. Indeed the good agreement between numerical
 simulations and experiments confirms the ability of our apparatus to
 measure {\it accurately} the statistics of partially coherent waves 
with broad spectra.  We have estimated from annex pulsed experiments that  temporal fluctuations as short as $250$fs can be precisely  measured with our
setup (see further publication). Our method enables for
example statistical study of fiber lasers  which are remarkable systems for the investigation of optical turbulence \cite{Turitsyna:13}.

Note that deviations from gaussian statistics have been reported
in 1D ``spatial experiments'' in which the transverse
intensity profile of optical
beams randomly fluctuating with space is easily recorded by using cameras
\cite{Bromberg:10,Onorato:13}. In spatial experiments, the speckle 
fields are localized \cite{Bromberg:10,Derevyanko:12}  and  the
dynamics of random waves is {\it confined} in a way comparable to the one found
in pulsed experiments \cite{Solli:12}. Our experiments with ``non-decaying''
continuous waves widen the perspective by providing new information
about nonlinear interactions among {\it unbounded}  random waves in
integrable turbulence.

Our experiment made in the anomalous dispersion regime is
qualitatively comparable to a one-dimensional deep-water-wave
experiment. Starting from random initial conditions, these hydrodynamical
experiments have demonstrated the formation of heavy-tailed 
statistics \cite{Onorato:04,Onorato:05}. However, in water-wave
experiments, the extreme events are far less frequent because the
height of the waves is strongly limited by the phenomenon of 
wave breaking \cite{Iafrati:13}. In hydrodynamics, relatively 
small deviations from gaussianity have  been observed and
interpreted in the framework of wave turbulence theory  
\cite{Nobuhito11,Janssen:03,Onorato:13}. On the contrary, our optical
fiber setup provides an accurate laboratory for the exploration  of
 strongly nonlinear  random wave systems ruled by the  1D-NLSE.\\

In this letter, we have quantified  how the
statistics of stochastic  nonlinear optical fields strongly deviate
from the normal law in an  optical fiber experiment. Our experiments
are very well described by the integrable focusing 1D-NLSE and they prove that
RW can appear in integrable turbulence \cite{Zakharov:13,
  Agafontsev:14,Agafontsev:14c}. Our experimental and numerical results
strengthen the idea that the emergence  
of deterministic solutions of 1D-NLSE such as AB, Peregrine solitons 
or KMs in nonlinear random fields is a  major mechanism for 
the formation of rogue waves \cite{Dudley:09,Akhmediev:13}. 
From our work, we hope to stimulate further theoretical and
experimental investigations permitting to establish a clear connection
between the emergence of coherent structures and the statistical
properties of nonlinear stochastic  fields.\\

This work was supported by the Labex CEMPI  (ANR-11-LABX-0007-01)
and by the French National Research Agency (ANR-12-BS04-0011 OPTIROC). 
The authors acknowledge S. Bielawski for fruitful discussions. The Authors acknowledge R. Habert, C. Szwaj and E. Roussel for technical help. The authors acknowlege Draka France compagny which has provided the fiber.

\end{document}